\documentclass[aps,pra,10pt,showpacs,tightenlines,notitlepage,nofootinbib,superscriptaddress,twocolumn,floatfix]{revtex4-2}

\usepackage{amsmath,amsfonts,amssymb,graphicx}
\usepackage[utf8]{inputenc}
\usepackage{bbold}
\usepackage{color,tikz}

\newcommand{\arXiv}[1]{\href{https://arxiv.org/abs/#1}{arXiv:#1}}

\begin{document}
\title{BQP $=$ PSPACE}
\author{Shibdas Roy}
\email{roy.shibdas@gmail.com}
\affiliation{Center for Quantum Engineering, Research and Education (CQuERE), TCG CREST, Salt Lake, Kolkata 700091, India.}
\affiliation{Department of Physics and Astronomy, University of Florence, 50019 Sesto Fiorentino, Italy.}

\begin{abstract}
The complexity class $PSPACE$ includes all computational problems that can be solved by a classical computer with polynomial memory. All $PSPACE$ problems are known to be solvable by a quantum computer too with polynomial memory and are, thus, known to be in $BQPSPACE$. Here, we present a polynomial time quantum algorithm for a $PSPACE$-complete problem, implying that $PSPACE$ is equal to the class $BQP$ of all problems solvable by a quantum computer in polynomial time. In particular, we outline a $BQP$ algorithm for the $PSPACE$-complete problem of evaluating a full binary $NAND$ tree. An existing best of quadratic speedup is achieved using quantum walks for this problem, so that the complexity is still exponential in the problem size. By contrast, we achieve an exponential speedup for the problem, allowing for solving it in polynomial time. There are many real-world applications of our result, such as strategy games like chess or Go. As an example, in quantum sensing, the problem of quantum illumination, that is treated as that of channel discrimination, is $PSPACE$-complete. Our work implies that quantum channel discrimination, and so, quantum illumination, can be performed efficiently by a quantum computer.
\end{abstract}

\maketitle

\section{Introduction}
Computational problems are classified into various complexity classes, based on the best known computational resources required by them \cite{MSbook}. The complexity class $PSPACE$ (Polynomial Space) is the class of all computational problems that can be solved by a classical computer using polynomial memory \cite{NC}. This class is one of the larger complexity classes, since the classes $P$ (Polynomial-time) - problems solvable by a classical computer in polynomial time \cite{MSbook}, $NP$ (Non-deterministic Polynomial-time) - problems not all solvable, but verifiable, by a classical computer in polynomial time \cite{MSbook}, and $BQP$ (Bounded-error Quantum Polynomial-time) - problems solvable by a quantum computer in polynomial time with a bounded probability of error \cite{NC}, are known to lie within $PSPACE$. Note that the class $NPSPACE$ (Non-deterministic Polynomial Space) equals $PSPACE$, since a deterministic Turing machine can simulate a non-deterministic Turing machine without needing much more space, although it may use much more time \cite{AB}.

It is known that the class $BQPSPACE$ (Bounded-error Quantum Polynomial Space) - problems solvable by a quantum computer with polynomial space (memory), equals the class $PSPACE$ \cite{JW}. Moreover, the class $QIP$ (Quantum Interactive Proof) - problems solvable by a quantum interactive proof system, is known to equal the class $PSPACE$ \cite{JJUW}. In this work, we show that $PSPACE\subseteq BQP$, i.e.~all problems in $PSPACE$ can be solved by a quantum computer in polynomial time. Since it is known that $BQP\subseteq PSPACE$ \cite{NC,BV}, our result implies that we must have $BQP=PSPACE$. Since $PSPACE$ is known to contain the class $NP$ \cite{KTbook}, it implies that all problems in $NP$ are also solvable by a quantum computer in polynomial time. We present a $BQP$ algorithm for a $PSPACE$-complete problem, i.e.~one of the hardest of all problems in $PSPACE$, thereby rendering $BQP=PSPACE$. A $PSPACE$-complete problem is totally quantified Boolean formula satisfiability (TQBF) \cite{HHL,MSbook}. We consider evaluating a complete binary $NAND$ tree as in Ref.~\cite{FGG} in polynomial time, since any total quantified Boolean formula can be expressed in terms of universal $NAND$ gates. Thus, our result implies that all problems, that can be solved by a classical computer with polynomial memory, can be solved by a quantum computer, not only with polynomial memory (since $BQPSPACE=PSPACE$), but also in polynomial time (since we show $BQP=PSPACE$).

There are many real-world applications of our result, such as playing games of strategy like chess or Go \cite{SA08}, or finding the shortest path without a map (Canadian traveller problem) \cite{PY}, or determining whether a regular expression generates every string over its alphabet \cite{HBH}. All these problems can be solved by a quantum computer in polynomial time, owing to our result here.

\section{Results}
\begin{figure}[!t]
\centering
\includegraphics[width=\linewidth]{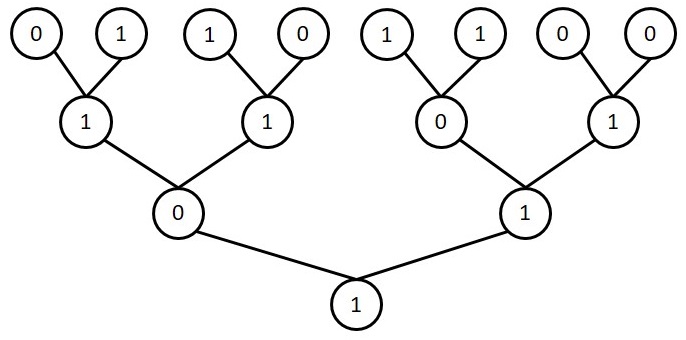}
\caption{A complete binary $NAND$ tree of depth $3$.}
\label{fig:nand}
\end{figure}

A complete binary $NAND$ tree has $N=2^n$ leaves, with values of either $0$ or $1$, assigned to each. The value of any other node is the $NAND$ of the two connected (adjacent) nodes just above (see an example in Figure \ref{fig:nand}). The goal of the algorithm is to evaluate the value at the root of the tree. Here, $n$ is the depth of the tree. While the classical randomized algorithm succeeds after evaluating $O(N^{0.753})$ of the leaves, the best known quantum algorithm succeeds in time $O(\sqrt{N})$, providing with a quadratic speedup \cite{FGG}.

In our approach, we consider $n$ qubits in a maximally-mixed state in a register $A$, that can express and index the $2^n$ leaves of the tree. We create a diagonal unitary $U$, that can suitably encode the $N$ leaf values into the phase factors of the eigenstates of the state in register $A$. Now, we need a superposition of the odd leaves only, and another superposition of the even leaves only, to perform phase estimation with of the unitary $U$. For this purpose, we denote the equal superposition of all the eigenstates of the state in register $A$, without the last qubit, as $|\chi\rangle$ and have this state in a register $B$. We perform phase estimation of the unitary $U$ for the eigenstates $|\chi\rangle|0\rangle$, using controlled-$U$ operations $cU$, to get the phase estimates of the odd leaves in a register $X$ entangled with $|\chi\rangle_B|0\rangle$. Next, we perform phase estimation of the unitary $U$ for the eigenstates $|\chi\rangle|1\rangle$ to get the phase estimates of the even leaves in a register $Y$. This is done by using register $Y$ as the phase estimate register, and the other register being $B$ and $|1\rangle$. So, after the phase estimations, we essentially have an entangled state in the three registers $X$, $Y$ and $B$. We, then, initialize another register $Z$ to $|1\rangle$, and perform the operation $CCNOT(X,Y,Z)$. This process ensures that the register $Z$ has the $NAND$ of the contents of the registers $X$ and $Y$. Register $B$ is now denoted as register $A$ for next iteration. We also perform quantum Fourier transform on register $Z$, to obtain the controlled-unitary $cU$ from register $ZA$ for next iteration. We repeat this process next of splitting the odd and even nodes, with register $B$ of $n-2$ qubits. This way the process is repeated a total of $\log_2(N)=n$ times. In the last iteration, register $Z$ will have the root value of the $NAND$ tree.

\subsection{Algorithm}
Our algorithm has the following steps:
\begin{enumerate}
\item \label{algo:step1} Given $N=2^n$ leaves of the tree, consider the following maximally-mixed state in a register $A$:
\begin{equation}\label{eq:state}
\psi = \frac{1}{2^n}\sum_{j=0}^{2^n-1}|j\rangle\langle j|.
\end{equation}
Create a diagonal unitary matrix that can encode the $N$ leaf values into the phase factors of the eigenstates of the state (\ref{eq:state}). The unitary is of the form:
\begin{equation}\label{eq:unitary}
U = \left[\begin{array}{cccccc}
e^{2\pi i\phi_0} & 0 &&& \ldots & 0\\
0 & e^{2\pi i\phi_1} &&& \ldots & 0\\
  &  & \cdot & & &\\
  &  & & \cdot & &\\
  &  & & & \cdot &\\
0 & 0 & & & \ldots & e^{2\pi i\phi_{N-1}}
\end{array}\right],
\end{equation}
where $\phi_j=u_j/2$. Here, $u_j=0$ if leaf value is $0$, and $u_j=1$ if leaf value is $1$. Set a variable $q=1$.

\item \label{algo:step3} Denote the equal superposition of the eigenstates of the state $\psi$ in register $A$, without the last qubit, as $|\chi\rangle$. Let us have this state in a new register $B$. For example, if $\psi := \frac{1}{4}\left(|00\rangle\langle 00|+|01\rangle\langle 01|+|10\rangle\langle 10|+|11\rangle\langle 11|\right)$, then we would have $|\chi\rangle := \frac{1}{\sqrt{2}}\left(|0\rangle+|1\rangle\right)$. Perform phase estimation of the unitary $U$ for the eigenstates $|\chi\rangle|0\rangle$ \cite{NC}, using controlled-$U$ operations $cU$, to obtain the phase estimates of the odd leaves in a register $X$ of $\ell=1$ qubit, entangled with $|\chi\rangle_B|0\rangle$. If $q<n$, the effective state in registers $X$ and $B$ is:
\begin{equation}
|\Psi\rangle = \frac{1}{\sqrt{2^{(n-q)}}}\sum_{j=0}^{2^{(n-q)}-1}|x_j\rangle|j\rangle,
\end{equation}
where $x_j=\tilde{\phi}_j$ in register $X$ for the odd leaves of the tree, $\tilde{\phi}_j$ being the estimate of the phase $\phi_j$. Here, register $B$ is of $n-q$ qubits. Next, perform phase estimation of the unitary $U$ for the eigenstates $|\chi\rangle|1\rangle$, using $cU$ operations, to obtain the phase estimates of the even leaves in a register $Y$, again of $\ell=1$ qubit. This is performed using register $Y$ as the phase estimate register, and the other register being $B$ and $|1\rangle$.

After the phase estimations for the odd and the even leaves of the tree, we effectively have three entangled registers $X$, $Y$ and $B$ in the following state, if $q<n$:
\begin{equation}
|\Phi\rangle = \frac{1}{\sqrt{2^{(n-q)}}}\sum_{j=0}^{2^{(n-q)}-1}|x_j\rangle|y_j\rangle|j\rangle,
\end{equation}
where $y_j=\tilde{\phi}_j$ in register $Y$ for the even leaves of the tree, $\tilde{\phi}_j$ being the estimate of the phase $\phi_j$.

\item \label{algo:step4} Add to $|\Phi\rangle$ another register $Z$ of $\ell=1$ qubit, initialised to $|1\rangle$:
\begin{equation}
|\Gamma\rangle = \frac{1}{\sqrt{2^{(n-q)}}}\sum_{j=0}^{2^{(n-q)}-1}|x_j\rangle|y_j\rangle|1\rangle|j\rangle.
\end{equation}

\item \label{algo:step5} Perform $CCNOT\left(X,Y,Z\right)$, so that register $Z$ has $NAND$ of registers $X$ and $Y$; for $q<n$, we get:
\begin{equation}
|\Omega\rangle = \frac{1}{\sqrt{2^{(n-q)}}}\sum_{j=0}^{2^{(n-q)}-1}|x_j\rangle|y_j\rangle|z_j\rangle|j\rangle,
\end{equation}
where $z_j = NAND(x_j, y_j) \, \forall j$. Notice that $|x_j\rangle$ (or $|y_j\rangle$) is $|0\rangle$ when the estimated phase is $0$, and $|1\rangle$ when the estimated phase is $1/2$ \cite{NC}. Denote register $B$ now as register $A$. That is, register $A$ is now of $n-q$ qubits. Also, denote the new maximally-mixed state in register $A$ as $\psi$.

\item \label{algo:step6} If $q<n$, perform quantum Fourier transform on register $Z$ in ${\rm Tr}_{XY}\left(|\Omega\rangle\right)$, to obtain the new controlled unitary $cU:|k\rangle|j\rangle\rightarrow|k\rangle e^{2\pi iz_jk/2}|j\rangle$. So, now we have $\phi_j=z_j/2$ in (\ref{eq:unitary}). If $q<n$, set $q=q+1$, and go back to step \ref{algo:step3}.

\item \label{algo:step7} At this step, register $A$ is empty, having no qubit. Measure register $Z$, and output the result as the desired value of the root of the $NAND$ tree.
\end{enumerate}

\subsection{Algorithm Complexity}
We analyse the complexity of our algorithm to demonstrate that it can be run on a quantum computer in polynomial, not exponential, time:
\begin{enumerate}
\item We need not create the initial state $\psi$ in register $A$, since it is not used anywhere. The $N$ number of leaf values of the $NAND$ tree can be encoded into the $N$ phase factors of the unitary $U$ at once. Since the unitary $U$, and so, the Hamiltonian $\mathcal{A}$, is a sparse matrix, $U=e^{i\mathcal{A}t}$ can be implemented efficiently in $O(\log_2(N)s^2t)=O(nt)$ steps \cite{BACS,HHL}, where $\mathcal{A}$ is an $(s=1)$-sparse matrix:
\begin{equation}
\mathcal{A} = \left[\begin{array}{cccccc}
2\pi\phi_0 & 0 &&& \ldots & 0\\
0 & 2\pi\phi_1 &&& \ldots & 0\\
  &  & \cdot & & &\\
  &  & & \cdot & &\\
  &  & & & \cdot &\\
0 & 0 & & & \ldots & 2\pi\phi_{N-1}
\end{array}\right].
\end{equation}
Thus, the complexity is $O(n)$, since $t\leq 2^\ell-1=1$.

\item The state $|\chi\rangle$ in register $B$ can be created easily using Hadamard gates, the complexity of which can be ignored. The phase estimation algorithm takes $O(\ell^2)=O(1)$ steps, mainly for computing the inverse Fourier transform \cite{NC}. Since we do phase estimation twice: once for odd leaves, and once for even leaves, the complexity of this step is $O(2\ell^2)=O(\ell^2)=O(1)$.

\item We can ignore the trivial complexity of this step.

\item A $3$-qubit Toffoli ($CCNOT$) gate requires $O(3)$ number of resources \cite{HLZWW} for its construction. The gate is required to be applied only once, and so, the complexity of this step is $O(3)$.

\item The quantum Fourier transform (QFT) has a complexity of $O(p^2)$ for $p$ qubits. In this step, we perform $n-q$ QFTs, but all in parallel. The complexity of each QFT is $O(\ell^2)=O(1)$, since the register $Z$ has $\ell=1$ qubit.

\item This step does not contribute to the complexity.
\end{enumerate}

Since all steps, except \ref{algo:step1} and \ref{algo:step7}, are repeated $n$ times, the overall complexity of our algorithm is $O(n)$ only.

\section{Discussion}
While classical algorithms for $PSPACE$-complete problems, such as totally quantified Boolean formula satisfiability (TQBF), take $O(N)$ steps in the worst case, the best known quantum algorithm achieves a quadratic speedup over classical algorithms, by using quantum walks \cite{FGG}. This algorithm has a complexity of $O(\sqrt{N})=O(2^{n/2})$, which is still exponential in $n$. By contrast, we achieve exponential quantum speedup with our algorithm, since our algorithm takes only $O(n)$ steps. Clearly, our algorithm beats the known $O(\sqrt{N})$ lower bound of quantum query complexity of evaluating $NAND$ trees \cite{HS}. In particular, the quantum query complexity of $O(\sqrt{N})$ for {\it formula evaluation} is known to be optimal \cite{ACRSZ}, which is evidently beaten by our algorithm. Our algorithm, therefore, implies that we must have $PSPACE\subseteq BQP$. Our result, along with the known result $BQP\subseteq PSPACE$ \cite{NC,BV}, implies that we must have $BQP=PSPACE$.

For example, in quantum sensing, the problem of quantum illumination, is $PSPACE$-complete. This is because quantum illumination is treated as a problem of channel discrimination, which is complete for the class $QIP$, that is equivalent to the class $PSPACE$ \cite{YMZZ}. The problem of quantum channel discrimination, and so, quantum illumination, can be solved by quantum computers in time, polynomial in the problem size, owing to our result.

Our result also implies that any post-quantum cryptography (PQC) \cite{DJBTL} protocol cannot use an $NP$-complete or a $PSPACE$-complete problem for secure communications, that can be safe from attacks by quantum computers. PQC algorithms must use problems, such as those from the class $EXPTIME$ (Exponential Time), that are not in $PSPACE$, for ensuring guaranteed security against any potential threat from quantum computers.

\section{Conclusion}
To conclude, we presented here the first $BQP$-algorithm for a $PSPACE$-complete problem, i.e.~evaluating a complete binary $NAND$ tree, thereby, proving $BQP=PSPACE$. The existing best quantum speedup for such problems is quadratic, owing to the Farhi-Goldstone-Gutmann algorithm \cite{FGG}. By contrast, our algorithm achieves an exponential quantum speedup, thereby allowing for such problems to be solved in polynomial time. Our work ensures that many real-world computationally difficult problems can be solved efficiently in polynomial time by quantum computers, such as the $PSPACE$-complete problems of $n\times n$ chess or Go strategy games, quantum illumination and quantum channel discrimination.

\begin{acknowledgments}
The author thanks Sreetama Das and Akshaya Jayashankar for useful discussions on this work. This work was financially supported by the European Union's Horizon 2020 research and innovation programme under FET-OPEN Grant Agreement No.~828946 (PATHOS).
\end{acknowledgments}

\bibliography{pspacebib}
\end{document}